\documentclass{emulateapj}

\lefthead{Bekki \& Chiba}
\righthead{Reionization effects on stellar halos}

\begin{document}

%\title{Searching for fossil records  of reionization by 
%wide-field imaging of galactic stellar halos}
\title{Fossil records of cosmic reionization in galactic stellar halos}

\author{Kenji Bekki} 
\affil{
School of Physics, University of New South Wales, Sydney 2052, Australia}

\and

\author{Masashi Chiba}
\affil{
Astronomical Institute, Tohoku University, Sendai, 980-8578, Japan\\}

\begin{abstract}

Galactic stellar halos have long been considered to contain
fossil information on early dynamical and chemical evolution of galaxies.
We propose that the surface brightness distributions of old stellar halos
contain the influence of reionization on early formation histories of galaxies.
By assuming that reionization significantly suppresses star formation
in small subgalactic clumps virialized after reionization
redshift ($z_{\rm reion}$), we first  numerically investigate 
how structural and kinematical properties  of
stellar halos formed from merging of these subgalactic clumps
depend on $z_{\rm reion}$.
We then discuss what observable properties of galactic stellar halos
offer us the fossil records of reionization influence on hierarchical formation
of halos based on the current results of numerical simulations.
We particularly suggest that both the half-light radius of stellar halos
and the slope of their surface brightness profile contain useful
information on when star formation in subgalactic clumps 
were significantly influenced by reionization.
By using the simulated surface brightness distributions of 
galactic stellar halos for models with different $z_{\rm reion}$,
we also discuss how wide-field imaging studies of extragalactic halos
will help us to elucidate the epoch  of cosmic reionization.
\end{abstract}


\section{Introduction}

Structural, kinematical, and  chemical properties of old,  metal-poor
stellar halos of galaxies have long provided valuable information on
early formation histories of galaxies, in particular,
galaxies in the Local Group (LG) (e.g., Freeman 1987; Majewski 1993;
Helmi et al. 1999; Morrison et al. 2000; Ibata et al. 2001). 
For example, the observed 
correlation between chemical and dynamical properties
of Galactic halo stars or lack thereof have enabled authors
to discuss whether the Galaxy was formed by a monolithic
collapse within a few dynamical timescales
(Eggen, Lynden-Bell, \& Sandage 1962; Norris \& Ryan 1991; 
Chiba \& Yoshii 1998).
Numerical simulations of Galaxy formation
based on the cold dark matter (CDM) model (e.g., Bekki \& Chiba 2000, 2001)
have demonstrated
that basic physical processes involved in the formation of the metal-poor, 
old stellar halo in a disk galaxy
are described by both dissipative and dissipationless merging of subgalactic
clumps and their resultant tidal disruption in the course
of gravitational contraction of the galaxy at high redshift:
Understanding of star formation and chemical evolution histories
in subgalactic clumps (i.e., progenitors of low-mass dwarfs) are key for
better understanding of the origin of galactic stellar halos.

Recently, several authors have extensively investigated
the possible physical  effects of reionization 
on galaxy formation, in particular, 
on star formation histories and
physical properties of old stellar populations in low-mass dwarfs 
(e.g.,  Bullock et al 2000; Gnedin 2000;
Susa \& Umemura 2004; 
Grebel \& Gallagher 2004; Willman et al. 2004). 
Some of these studies  suggested that 
suppression of star formation 
resulting from reionization can be a very important
physical process for better understanding the observability
and the early star formation activities in dwarfs.
These previous  results, combined with those on stellar halo formation
via hierarchical merging of low-mass subgalactic clumps
(Bekki \& Chiba 2000; 2001),
therefore imply that 
%{\it if reionization has significant influence
%on the formation of old dwarf galaxies,
%such influence can be reflected on physical properties of
%the old stellar halos in luminous galaxies.}
{\it the influence of reionization on the formation of dwarf galaxies may be
imprinted in the physical properties of old stellar halos in bright galaxies.}

Although many observational and theoretical studies in the {\it young} universe
have already provided important constraints on
the cosmic reionization history (e.g., Gunn \& Peterson 1965;  
Benson et al. 2001; Fan et al. 2003; Santos 2004; Wyithe \& Loeb 2004),
it remains yet to be explored in the {\it present-day} universe as to
(1) whether old stellar halos of galaxies contain the fossil records of
reionization history and
(2) what observations of stellar halos  can provide what clues to
the possible influences of reionization on galaxy formation processes.
Physical properties of metal-poor, old stellar halos 
in galaxies beyond the LG have been recently investigated by 
the {\it Hubble Space Telescope} ({\it HST})
and large ground--based telescopes (e.g., Harris et al. 2000; 
Gregg et al. 2004).
Furthermore, recent numerical studies provided some specific predictions
on star formation histories of dwarf galaxies under
the influence of reionization (e.g., Susa \& Umemura 2004).
Thus it is quite timely to discuss the above questions based
on the comparison between observations and
numerical simulations addressing the possible influence of reionization
on the properties of galactic stellar halos.

The purpose of this Letter is to show 
that wide-field imaging of galactic halos
by large ground-based telescopes (e.g., Suprime-Cam on Subaru)
can provide vital clues to the possible influences of reionization
on early formation history of galaxies.
We first present the results of our numerical simulations and stress
how structural and kinematical properties  of
stellar halos, which are formed from merging of small subgalactic clumps,
depend on the epoch of reionization ($z_{\rm reion}$).
We then discuss  (1) the origin of diverse stellar halo properties
of galaxies of the LG  in terms of reionization effects on halo formation
and (2) the importance of ongoing and future
wide-field optical imaging of the halos in better understanding
galaxy formation.
Possible reionization effects on structural properties of  
globular cluster systems
in disk galaxies are discussed in Bekki (2005).

\section{The model}

We simulate the formation of a Milky Way sized galaxy halo
in a $\Lambda$CDM Universe with ${\Omega} =0.3$, 
$\Lambda=0.7$, $H_{0}=70$ km $\rm s^{-1}$ ${\rm Mpc}^{-1}$,
and ${\sigma}_{8}=0.9$,
and thereby investigate merging/accretion
histories of subhalos that can contain low mass dwarfs.
The way to set up initial conditions for the numerical simulations
is essentially
the same as that adopted by Katz \& Gunn (1991) and Steinmetz \& M\"uller
(1995). 
We consider an isolated homogeneous, rigidly rotating sphere, on which
small-scale fluctuations according to a CDM power spectrum are superimposed.
The initial total mass ($M_{\rm t}$), radius, spin parameter ($\lambda$),
and the initial overdensity ${\delta}_{i}$   of the sphere
in the standard  model are $6.0\times10^{11}\rm M_{\odot}$, 30 kpc, 
0.08, and 0.26, respectively. 
Initial conditions similar to those adopted in the present study
are demonstrated to be plausible and realistic for the formation
of the Galaxy (e.g., Steinmetz \& M\"uller 1995; Bekki \& Chiba 2001).
In addition to this standard models, we investigate models
with different $M_{\rm t}$ and ${\delta}_{i}$.

We start the collisionless simulation at $z_{\rm start}$ (=30) and follow it 
till $z_{\rm end}$ (=1) to identify virialized subhalos
with the densities larger than $170 {\rho}_{\rm c}(z)$,
where ${\rho}_{\rm c}(z)$ is the critical density of the universe, 
at a redshift $z$.
This $170 {\rho}_{\rm c}(z)$ corresponds to the mean mass denstiy
of a collapsed and thus gravitationally bound
object at $z$ (e.g., Padmanabhan 1993).
The minimum number of particles within a virialized subhalo
($N_{\rm min}$) is set to be 32 corresponding to the mass resolution
of 3.8 $\times$ $10^{7}$ $M_{\odot}$. This number of 32 is chosen
so that we can find a virialized object at a given $z$ in a robust manner.
For each individual virialized subhalo
with the virialized redshift of $z_{\rm vir}$,
we estimate a radius ($r_{\rm b}$) within which 20 \%  of the total mass
is included,  and then the particles within $r_{\rm b}$ are labeled 
as ``baryonic'' particles. This procedure for defining baryonic particles
is based on the assumption that energy dissipation via radiative cooling
allows baryon to fall into the deepest potential well of dark-matter halos.
Such baryonic particles in a subhalo will be regarded as candidate
``stellar'' particles to form stellar halos in the later dynamical stage,
if the subhalo is later destroyed and baryonic particles initially within
the subhalo is dispersed into the galactic halo region.
Thus, the present dissipationless models track the formation of stellar halos
via hierarchical merging of subhalos, although the models are not adequate
to the study of star formation histories in subhalos (as was done
in our previous studies, e.g., Bekki \& Chiba 2001).

Previous theoretical studies have demonstrated that 
ultraviolet background radiation in a reionized
universe can significantly reduce the total amount of cold HI 
and molecular gas 
that are observed to be indispensable for galactic active star formation
(e.g., Susa \& Umemura 2004).
In order to investigate this suppression effects of star formation
on the final structural
properties of the simulated stellar halos,
we adopt the following idealized assumption:
{\it If a subhalo is virialized after the completion
of the reionization ($z_{\rm reion}$),
star formation is totally suppressed in such a subhalo.}
Then, hypothetical baryonic particles in the
subhalos with $z_{\rm vir}$ $<$ $z_{\rm reion}$ will {\it not} be
identified as stellar halo particles in the later stage, but
those in the subhalos with $z_{\rm vir}$ $\ge$ $z_{\rm reion}$
will be regarded as progenitors of visible stellar halos.
Recent WMAP ({\it Wilkinson Microwave Anisotropy Probe})
observations have shown that plausible $z_{\rm reion}$  
ranges from 11 to 30 (Spergel et al. 2003; Kogut et al. 2003)
whereas quasar absorption line studies give the lower limit
of 6.4 for $z_{\rm reion}$ (Fan et al. 2003).
Guided by these observations, we investigate the models
with $z_{\rm reion}$  = 0 (no reionization), 6, 8, 10, 13, and 15.
The  adopted picture of single  epoch of reionization might well be
somewhat oversimplified and less realistic (e.g., Gnedin 2004),
however, this idealized model can help  us to elucidate some essential
ingredients of the reionization effects on stellar halo formation.

All the calculations have been carried out on the GRAPE board
(Sugimoto et al. 1990).
Total number of particles used in our simulations is 508686 
and the gravitational softening length is 0.38 kpc.
%We consider that the final structure of the simulated
%stellar halo at $z_{\rm end}$ (=1) is the same as that at $z$ = 0,
%because the halo is dynamically relaxed completely  until $z_{\rm end}$ 
We note that later accretion of satellites at $z < 1.5$ is minor
in the final structures of the simulated halos at $z = 0$, so
the calculation is ended at $z_{\rm end} = 1$ to obtain the dynamically
relaxed halo structures.
We used the COSMICS (Cosmological Initial Conditions and
Microwave Anisotropy Codes), which is a package
of fortran programs for generating Gaussian random initial
conditions for nonlinear structure formation simulations
(Bertschinger 1995). 
In order to derive 2D distributions of 
$B-$band surface brightness (${\mu}_{\rm B}$) from  the simulated
density distributions of stellar halos, 
we assume that the stellar mass-to-light-ratio
of 4.0 that is a reasonable value for stars with ages of $\sim 13$ Gyr
and metallicities of [Fe/H] $\sim$ $-1.6$ (Vazdekis et al. 1996).

\section{Result}

Figure 1 shows that the projected radial density profiles
of stellar halos depend strongly on $z_{\rm reion}$
in such a way that they are steeper 
for the models with higher $z_{\rm reion}$.
This result implies that radial density profiles of 
galactic stellar halos can contain fossil information on
the reionization epoch.
The physical reason for this dependence is described as follows.
For the model with higher $z_{\rm reion}$ (=15),
only subgalactic clumps  that are virialized before $z_{\rm reion}$ 
can have old stars (i.e., no suppression of star formation by reionization)
within it. They merge with one another in the early stage of the galactic
collapse to form the first, single massive stellar halo around $z \sim 2$
with a steep density profile resulting from violent relaxation of the merging.
After that,  subgalactic clumps that are virialized later than $z_{\rm reion}$
and thus do not contain stellar components merge with and accreted by
the first massive halo. 
These lately merging subgalactic clumps are highly likely to
be destroyed in the outer part of the galaxy.
Consequently, if these subgalactic clumps contain
stars initially within them,
they are  dispersed into the outer halo region
and thus significantly flatten the radial density profile of the stellar halo. 
For the model with high $z_{\rm reion}$ (=15), such later accretion
cannot flatten the radial density profile of the stellar halo so that
it can show a steeper density profile. 
Structural evolutoin of dark-matter halos due to hierarchical clustering
process has already been demonstrated by Zhao et al. (2003), and the same
process is at work for the $z_{\rm reion}$-dependence of stellar halos
in the present study.

Figure 2 shows how surface brightness (${\mu}_{\rm B}$) distributions 
of galactic stellar halos in the $B$ band
depend on $z_{\rm reion}$.
It is clear from Figure 2 that the model with higher $z_{\rm reion}$ (=15)
shows a sharp edge in its projected light distribution of the stellar halo
in comparison with the model with lower $z_{\rm reion}$ (=6).
The ratio of the half-mass radius of the stellar halo
% ($R_{\rm h,s}$)
to that of the dark matter halo
% ($R_{\rm h,d}$)
is 0.15 for 
$z_{\rm reion}=15$ and 0.71 for $z_{\rm reion}=6$, which suggests that
$z_{\rm reion}$ can control the compactness of the stellar halo distribution
of a galaxy. Figure 2 demonstrates that the outer stellar halo 
($10 < R < 20$~kpc)  
is much fainter in the model with  $z_{\rm reion}=15$ 
(${\mu}_{\rm B}$ $\sim$  $28-30$ mag arcsec$^{-2}$) than in
that with $z_{\rm reion}$ = 6 (${\mu}_{\rm B}$ $\sim$  
$25-27$ mag arcsec$^{-2}$).

Figure 2 furthermore shows that
a few stellar substructures can be seen in the outer stellar halo only for
the model with higher $z_{\rm reion}$ (=15),
which implies that the observability of stellar halo substructures
(or inhomogeneity in luminosity distribution) can also depend on $z_{\rm reion}$. 
Irrespective of $z_{\rm reion}$, the stellar halos have prolate-triaxial shapes
that follow the shapes of the dark matter halos in the present dissipationless
simulations. Therefore, the ${\mu}_{\rm B}$  distributions of stellar halos depend 
on from which direction we observed the halos.
It is accordingly  noted here that the differences in shapes of stellar halos
seen in Figure 2 are  not due to $z_{\rm reion}$. 
Thus the way of radially falling ${\mu}_{\rm B}$  of stellar halos 
(i.e., presence or non-presence of outer sharp edges in ${\mu}_{\rm B}$),
compactness of the halos, 
and the average value of ${\mu}_{\rm B}$
can be regarded as fossil records
of $z_{\rm reion}$.

Figure 3 shows that ${\mu}_{\rm B}$ distributions
of galactic stellar halos depend on the initial mass densities of
galaxies (${\delta}_{\rm i}$) in the sense that
the galactic stellar halo in the model with larger ${\delta}_{\rm i}$ (=0.39),
shows higher ${\mu}_{\rm B}$ as a whole.  
This is because total number of subgalactic clumps that can be virialized
well before $z_{\rm reion}$ is larger in the model with larger ${\delta}_{\rm i}$.
This result implies that if reionization can strongly suppress
star formation for subgalactic clumps virialized after $z_{\rm reion}$, 
${\mu}_{\rm B}$ distributions of stellar halos  
can be remarkably diverse even for galaxies with similar masses 
(yet different  ${\delta}_{\rm i}$)
owing to the ${\delta}_{\rm i}$-dependence of the reionization effects.

Kinematical properties of stellar halos in the models with a given mass
depend weakly on $z_{\rm reion}$. 
For example,
the maximum rotational velocity $V_{\rm m}$ of stellar halos
is appreciably larger in the model with $z_{\rm reion}$ = 6
($V_{\rm m}\sim20$ km s$^{-1}$)
than that with  $z_{\rm reion}$ = 15  
($V_{\rm m}\sim10$ km s$^{-1}$).
This result of more rotation in the model with smaller $z_{\rm reion}$ 
is described as follows.
For the model with $z_{\rm reion}$ = 6,
subgalactic clumps which are virialized at 
$5 \le z \le 15$ and later merge with the galaxy 
contain a larger amount of orbital angular momentum.
Therefore,
the outer halo stars formed from tidal destruction of these clumps
can have a larger amount of intrinsic angular momentum after
conversion of orbital angular momentum into
intrinsic one in the merging processes.

\placefigure{fig-1}
\placefigure{fig-2}

\placefigure{fig-3}
\placefigure{fig-4}

\section{Discussions and conclusions}

 Previous observations suggested that 
the radial density profile  ($\rho(r)$) of the Galactic stellar halo
can be well described as $\rho(r) \propto r^{-3.5}$ 
corresponding roughly  to $\Sigma(R) \propto R^{-2.5}$ in the projected profiles 
(e.g., Sommer-Larsen \& Zhen 1990; Chiba \& Beers 2000).
Considering the present result that $\rho(r)$ of galactic stellar halos
in the  models with higher $z_{\rm reion}$ (=15) are as steep as 
$r^{-3.5}$
and thus consistent with the observations,
we suggest that the origin of the observed steep $\rho(r)$ could be closely associated
with reionization effects on the formation of subgalactic clumps of the Galaxy at
a relatively higher redshift ($\sim15$).
Furthermore,
the simulated halos with sharply truncated outer distributions in the models with
higher $z_{\rm reion}$
suggest that the observed apparently sharp edge in the M31 stellar halo
(e.g., Ferguson et al. 2002)
might well result from such reionization effects.

 One of the most remarkable results of previous observations
for the M31 stellar halo  is that
it is dominated by moderately high-metallicity 
([Fe/H] $\sim$ $-0.5$) populations and shows a high stellar density
compared with the Galactic stellar halo 
(e.g., Durrell, Harris, \& Pritchet 2001; Reitzel \&  Guhathakurta 2001). 
The origin of  the high metallicity, high-density stellar halo
of M31 was discussed in terms of the formation of M31's bulge via
past major merger event (Bekki et al. 2003). We here suggest 
an  alternative scenario 
that the observed differences
in metallicity distributions and structures of stellar halos
between the two that have similar masses (Evans et al. 2000)
can be  due to M31's higher 
initial densities by which star formation and chemical enrichment
proceeded more efficiently before reionization compared with
the Galaxy.
We also suggest that the physical reason
for the observed possible difference
in the mean metallicities of stellar halos between
M33 ([Fe/H] $\sim$ $-1.3$; Brooks et al. 2004)
and the LMC  ([Fe/H] $\sim$ $-1.5$; Borissova et al. 2004) 
might well be the same as above for the Galaxy-M31 difference.

Although observations suggest that structural properties and metallicity
distributions of galactic stellar halos  in the LG could be 
quite diverse (e.g., Durrell et al. 2001; Alves 2004; Brooks et al. 2004), 
it remains totally unclear whether such diversity
can be seen in other galaxies beyond the LG.
Our simulations showed that 
physical properties of stellar halos in galaxies can be diverse, 
partly because initial protogalactic conditions of a galaxy
(e.g., the mass density) can determine 
to what extent reionization can influence star formation histories of
subgalactic clumps that form the galaxy.
In order to reveal radial structures of stellar halos in galaxies
beyond the LG and possible influences of reionization on the halo formation 
in these galaxies,
we need to conduct wide-field imaging for {\it global}
light distributions in galaxies and measure
the surface brightness distributions of the halos down to $28-30$ mag arcsec$^{-2}$.

For example, if we try to cover the 60kpc$\times$60kpc halo region
for a galaxy with the distance of $\sim$ 10 Mpc,
we need $20^{'} \times 20^{'}$ wide-field imaging by a large telescope
than can detect low-surface brightness stellar halos. 
Such wide-field imaging naturally enables us to reveal 
some stellar halo properties possibly influenced by reionization,
such as the outer  sharp edges in the surface brightness distributions
and the steep radial luminosity profiles.
Statistical studies of stellar halo properties by wide-field imaging 
for galaxies in different environments (e.g., clusters vs fields)
can also help us to understand different degrees of reionization effects
on the stellar halo formation in different environments, 
which could result from possible inhomogeneity in reionization processes 
(e.g., Benson et al. 2001; Wyithe \& Loeb 2004).
This type of observations might well be formidable tasks for
8m-telescopes with wide-field imaging facilities: for instance,
a few hour exposure with Subaru Suprime-Cam would reach as deep as
29 mag arcsec$^{-2}$ in the $B$-band with sufficient S/N ratios.
Systematic observational studies of stellar halos by
such telescopes will enable us to discuss differences in reionization
influences on stellar halo formation between different galaxies
and thus to understand better the origin of galactic stellar halos.

\acknowledgments
We are  grateful to the anonymous referee for valuable comments,
which contribute to improve the present paper.
K.B. acknowledges the financial support of the Australian Research 
Council throughout the course of this work.
The numerical simulations reported here were carried out on GRAPE
systems kindly made available by the Astronomical Data Analysis
Center (ADAC) at National Astronomical Observatory of Japan (NAOJ).

\clearpage

%%%%%%%%%%%%%%%%%%%%%%% Figure Captions

\newpage
\plotone{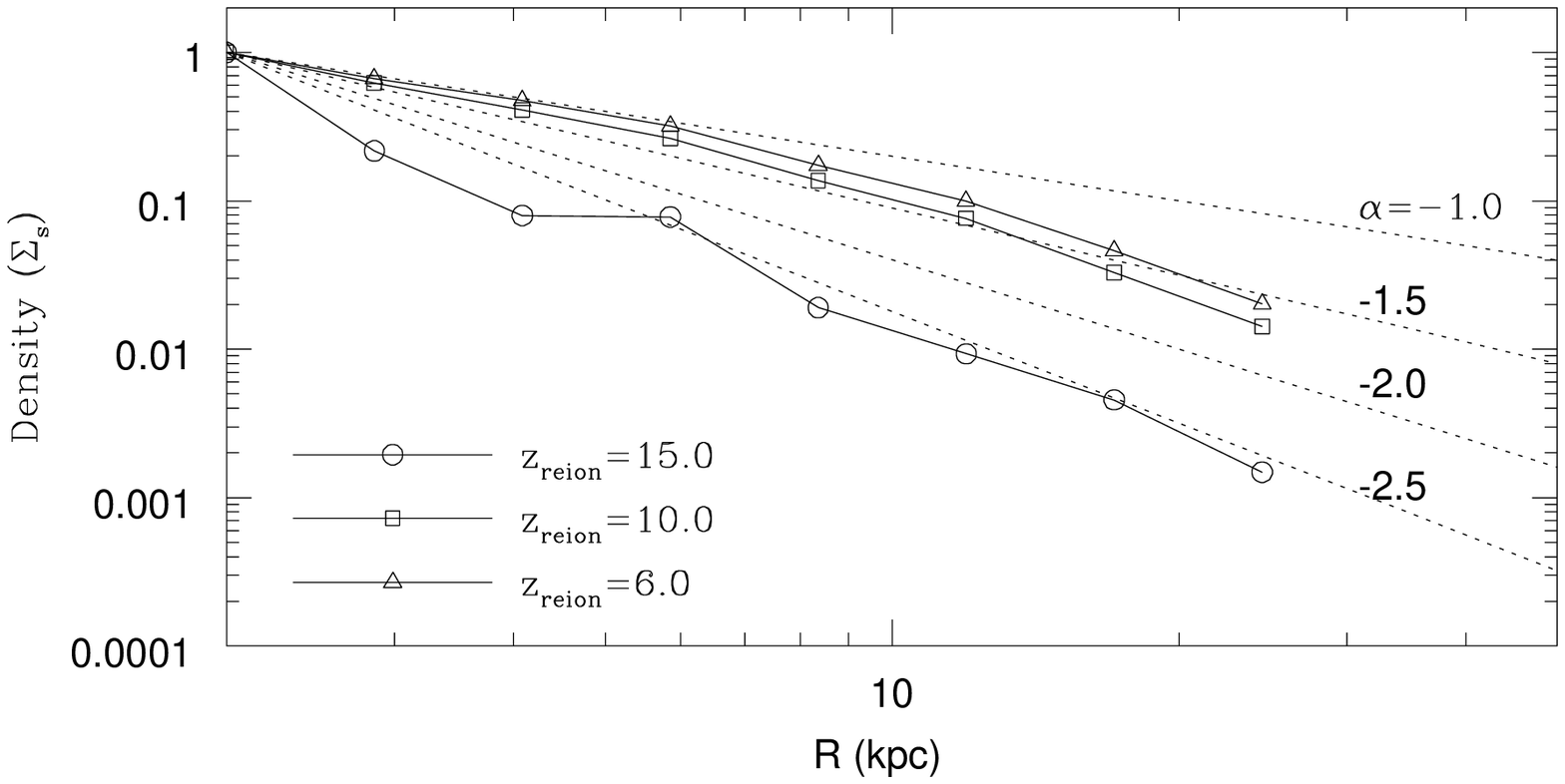}
\figcaption{
Projected radial density profiles (${\Sigma}_{\rm s}(R)$) of stellar halos
for the standard models with $z_{\rm reion}=15$ (open circles),
$z_{\rm reion}=10$ (open square),
and $z_{\rm reion}=6$ (open triangle).
For clarity, ${\Sigma}_{\rm s}(R)$ normalized to the central value
in each model is shown.
For comparison, the power-law density profiles with 
${\Sigma}_{\rm s}(R) \propto R^{\alpha}$ for 
$\alpha$ = $-1.0$, $-1.5$, $-2.0$, and $-2.5$ are 
shown by dotted lines.
Note that models with higher $z_{\rm reion}$ show steeper 
profiles.
The apparent bump at $R \sim 6$kpc for  $z_{\rm reion}=15$
is due to a stellar substructure.
\label{fig-1}}

\newpage
\plotone{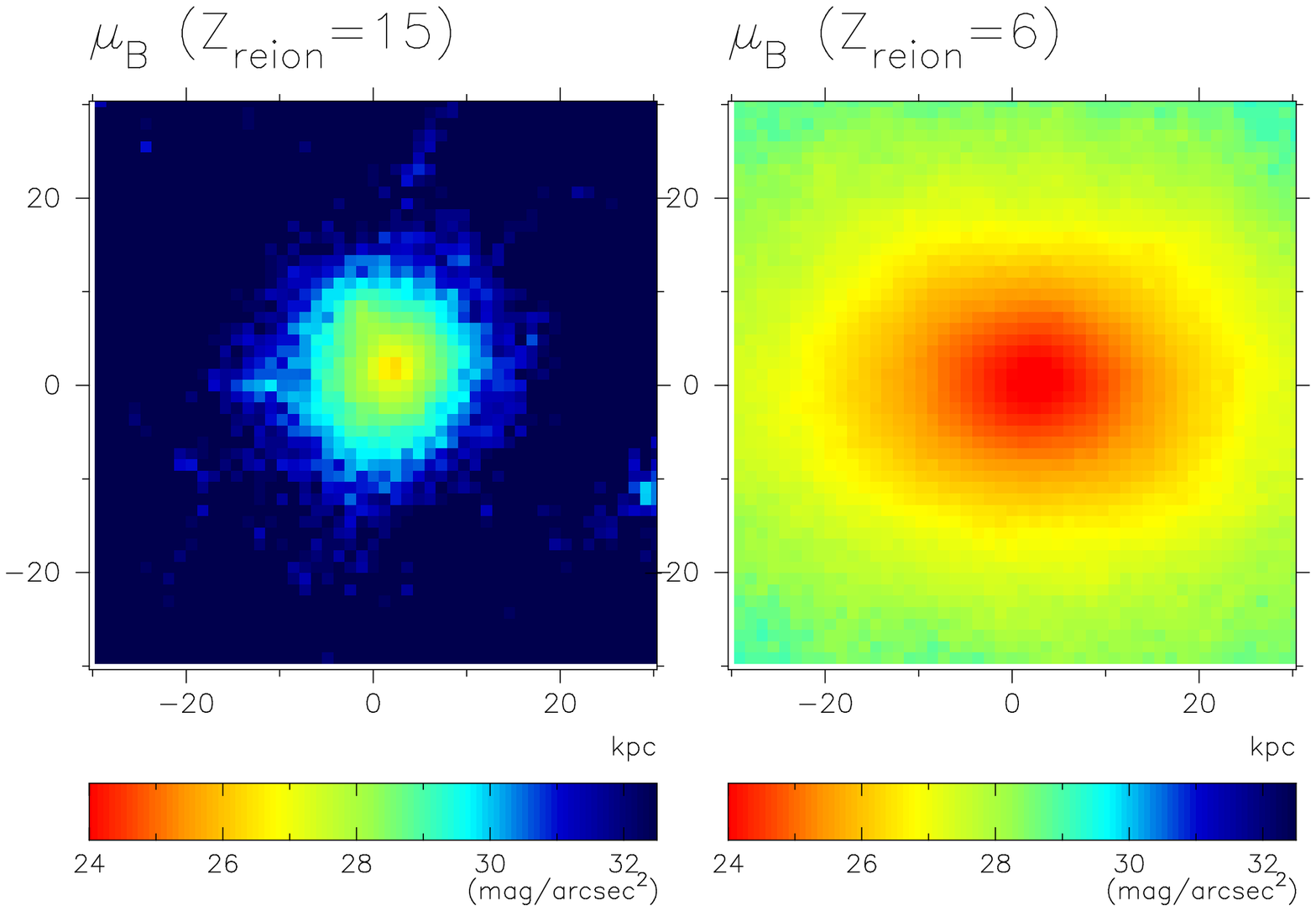}
\figcaption{
The 2D distribution of surface brightness ${\mu}_{\rm B}$ (mag arcsec$^{-2}$) 
for the two extreme cases with $z_{\rm reion}$=15 (left)
and $z_{\rm reion}$=6 (right) for the standard models.  
This figure illustrates how $z_{\rm reion}$ affects
the observed ${\mu}_{\rm B}$ distribution of
stellar halos.
To derive these ${\mu}_{\rm B}$ distributions,
we divide the $60$kpc $\times$ $60$kpc halo region 
of a simulation
into 50 $\times$ 50 meshes
and thereby estimate ${\mu}_{\rm B}$ for each mesh
according to the photometric model described
in the main text.
We also smooth out the 2D distributions by using
a Gaussian kernel with the smoothing length of 6 kpc.
\label{fig-2}}

\newpage
\plotone{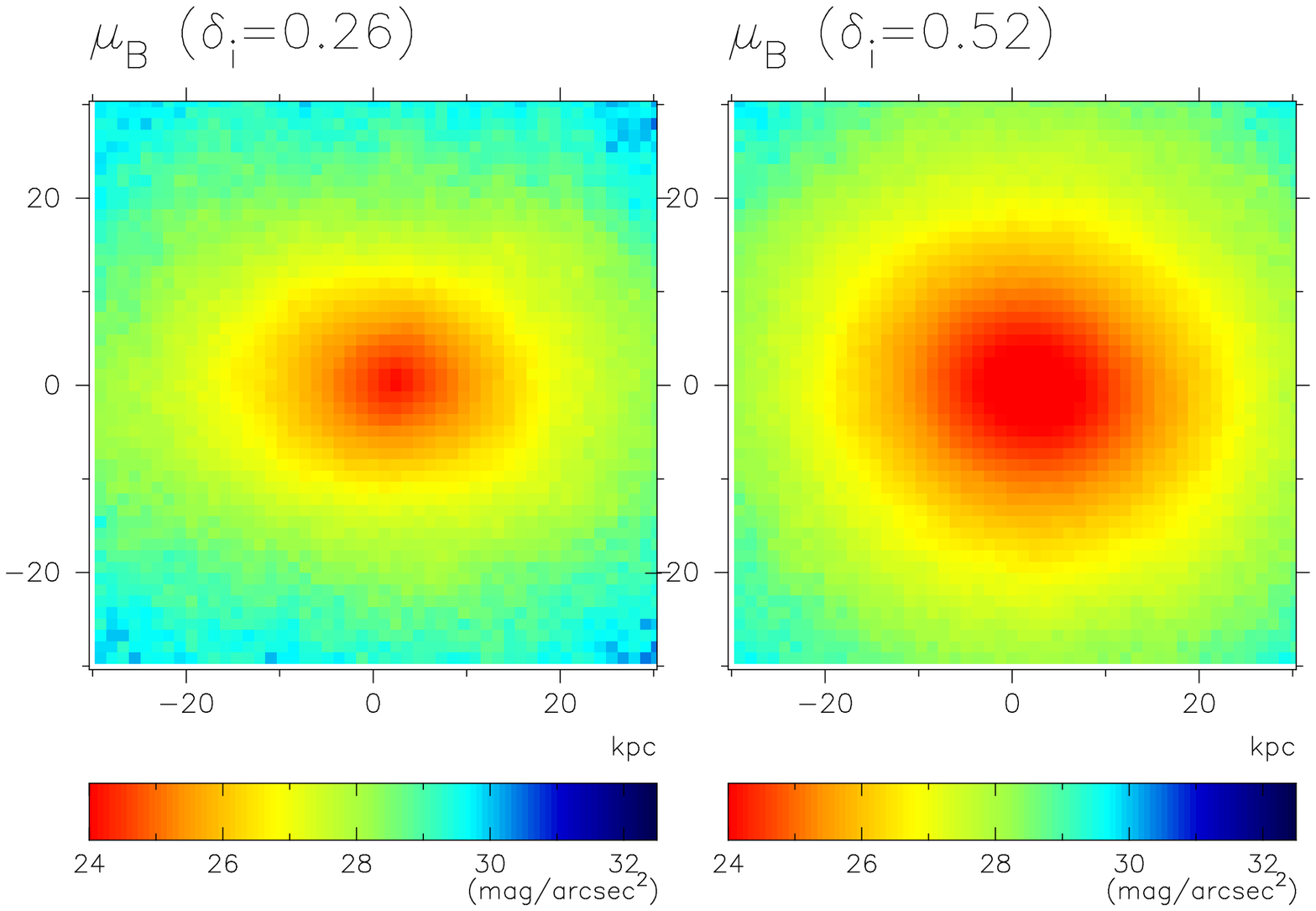}
\figcaption{
The same as Figure 3 but for the two models with
the same $z_{\rm reion}$ (=10) yet different
initial over denisites:
${\delta}_{\rm i}=0.26$ (left) 
and ${\delta}_{\rm i}=0.52$ (right).
\label{fig-3}}
\end{document}